\documentclass[prd,preprintnumbers,nofootinbib,preprint]{revtex4} 
\usepackage{graphicx} 
\usepackage{amsmath}
\usepackage{amsfonts,amsbsy}
\usepackage{amssymb}
\usepackage{subfig}

\def\gsim{ \,\, \vcenter{\hbox{$\buildrel{\displaystyle >}\over\sim$}}
 \,\,}

\def\be{\begin{equation}}
\def\ee{\end{equation}}
\def\bea{\begin{eqnarray}}
\def\eea{\end{eqnarray}}

\begin{document}

\title{Initial conditions for dipole evolution
  beyond the McLerran-Venugopalan model}

\preprint{RBRC-935}

\author{Adrian Dumitru$^{a,b,c}$ and Elena Petreska$^{b,c}$}
\affiliation{
$^a$ RIKEN BNL Research Center, Brookhaven National
  Laboratory, Upton, NY 11973, USA\\
$^b$ Department of Natural Sciences, Baruch College, CUNY,
17 Lexington Avenue, New York, NY 10010, USA\\
$^c$ The Graduate School and University Center, City
  University of New York, 365 Fifth Avenue, New York, NY 10016, USA}

\begin{abstract}
We derive the scattering amplitude $N(r)$ for a QCD dipole on a dense
target in the semi-classical approximation. We include the first
subleading correction in the target thickness arising from $\sim
\rho^4$ operators in the effective action for the large-$x$ valence
charges. Our result for $N(r)$ can be matched to a phenomenological
proton fit by Albacete {\it et al.\ } over a broad range of dipole
sizes $r$ and provides a definite prediction for the $A$-dependence
for heavy-ion targets. We find a suppression of $N(r)$ for finite $A$
for dipole sizes a few times smaller than the inverse saturation
scale, corresponding to a suppression of the classical {\it
  bremsstrahlung} tail.
\end{abstract}


\maketitle

In this paper we derive the dipole scattering amplitude
$N(r)$~\cite{Mueller:1993rr}
on a dense target in the semi-classical approximation. We
restrict to a perturbative expansion of the Wilson lines valid at
short distances $r$ but include the first subleading (in density)
correction arising from $\sim \rho^4$ operators in the effective
action for the large-$x$ valence charges.

Our result may prove useful for a better theoretical understanding of
the $A$-dependence of the initial condition for high-energy evolution
of $N(r)$, as discussed in more detail below.

Let us first recall the setup for the McLerran-Venugopalan (MV)
model~\cite{MV}. The ``valence'' charges with large light-cone
momentum fraction $x$ are described as recoilless sources on the
light cone with $\rho^a(x_\perp,x^-)$ the classical color charge
density per unit transverse area and longitudinal length
$x^-$. Kinetic terms for $\rho$ are neglected since transverse momenta
are assumed to be small. In the limit of a very high
density of charge $\rho$ the fluctuations of the color charge, by the
central limit theorem, are described by a Gaussian effective action
\bea 
S_{MV}[\rho] = \int d^2\bold x_\perp\int^\infty_{-\infty} dx^-~
\frac{\rho^a(x^-,\bold x_\perp) \, \rho^a(x^-,\bold
  x_\perp)}{2\mu^2(x^-)}~. \label{eq:S_MV}
\eea
Here, $\mu^2(x^-)dx^-$ is the density of color sources per unit
transverse area in the longitudinal slice between $x^-$ and $x^-+dx^-$
and
\be
\int_{-\infty}^\infty dx^-\mu^2(x^-) \sim g^2A^{1/3}
\ee
is proportional to the thickness $\sim A^{1/3}$ of the target
nucleus. This action can be used to evaluate the expectation value of
the dipole operator
\bea
D(r) &\equiv& \frac{1}{N_c}\langle \mathrm{tr}~V(\bold x_\perp)
V^\dagger (\bold y_\perp) \rangle  \\
&=& \exp\left(-\frac{g^4 C_F}{8\pi} \int dx^- \mu^2(x^-)~r^2
\log\frac{1}{r\Lambda}\right) \\
&=& \exp\left(-\frac{Q_s^2\,r^2}{4} \log\frac{1}{r\Lambda}\right)~,
\label{eq:N_MV}
\eea
where $r\equiv |\bold x_\perp-\bold y_\perp|$ and $\Lambda$ is an
infrared cutoff on the order of the inverse nucleon radius; the
explicit $r$-dependence has been obtained in the limit $\log
1/(r\Lambda)\gg1$. The scale $Q_s$ denotes the saturation momentum at
the rapidity of the sources. For details of the calculation we refer to
refs.~\cite{JalilianMarian:1996xn,hep-ph/9802440,GelisPeshier} and to
the appendix below.

To go beyond the limit of infinite valence charge density one
considers a ``random walk'' in the space of SU(3) representations
constructed from the direct product of a large number of fundamental
charges~\cite{JV}. The effective action describing color fluctuations
then involves a sum of the quadratic, cubic, and
quartic Casimirs which can be written in the form~\cite{arXiv:1105.4155}
\bea
S[\rho] = \int d^2\bold v_\perp\int^\infty_{-\infty} dv_1^- \left\{
\frac{\rho^a(v_1^-,\bold v_\perp) \rho^a(v_1^-,\bold
  v_\perp)}{2\mu^2(v_1^-)}-\frac{d^{abc}\rho^a(v_1^-,\bold
  v_\perp)\rho^b(v_1^-,\bold v_\perp)\rho^c(v_1^-,\bold
  v_\perp)}{\kappa_3} \nonumber \right.\\ 
\left. +\int^\infty_{-\infty}dv_2^-\, \frac{\rho^a(v_1^-,\bold
  v_\perp)\rho^a(v_1^-,\bold v_\perp)\rho^b(v_2^-,\bold
  v_\perp)\rho^b(v_2^-,\bold v_\perp)}{\kappa_4}\right\}~.
\label{eq:Squartic}
\eea
The coefficients of the higher dimensional operators are
\bea
\kappa_3 &\sim & g^3A^{2/3}~, \\
\kappa_4 &\sim & g^4A~,
\eea
and so involve higher powers of $g A^{1/3}$. In what follows we
restrict to leading order in 1/$\kappa_4$ and drop the ``odderon''
operator $\sim \rho^3$ from~(\ref{eq:Squartic}) since it does not
contribute to the expectation value of the dipole operator at leading
order in $1/\kappa_3$. The details of the calculation are shown in the
appendix, here we just quote the final result for the $T$-matrix
\bea \label{eq:Nr_quartic} 
N(r) &\equiv& 1-D(r)   \nonumber\\
&=& \frac{Q_s^2 r^2}{4}\log\frac{1}{r \Lambda}-
\frac{C_F^2}{6\pi^3}\frac{g^8}{\kappa_4}
\left[\int_{-\infty}^\infty dz^- \mu^4(z^-)\right]^2 r^2\log^3\frac{1}{r \Lambda}
~~~,~~~(r^2 Q_s^2 < 1)~.
\eea
This now involves a new moment of the valence color charge
distribution, namely $\int dx^- \mu^4(x^-)$. We have calculated
the $\mathcal{O}(1/\kappa_4)$ correction analytically only in the ``short
distance'' regime up to order $\sim r^2$. Recall, however, that the
effective theory~(\ref{eq:S_MV}) or~(\ref{eq:Squartic}) does not apply
to the DGLAP regime at asymptotically short distances. Our result
could in principle be extended into the saturation region $r\gsim
1/Q_s$ by generating the color charge configurations $\rho^a(x^-,\bold
x_\perp)$ non-perturbatively, numerically~\cite{Lappi:2007ku}.

The dipole scattering amplitude for a proton target has been fitted in
ref.~\cite{AAMQS} to deep-inelastic scattering data. The
Albacete-Armesto-Milhano-Quiroga-Salgado (AAMQS) model
for the initial condition for small-$x$ evolution is given by
\be
N_{\rm AAMQS}(r,x_0=0.01) = 1 - \exp \left[- \frac{1}{4}
\left(r^2 Q_s^2(x_0) \right)^\gamma \log\left(e+\frac{1}{r \Lambda}\right)
  \right]~,  \label{eq:N_AAMQS}
\ee
with $\gamma\simeq1.119$ \footnote{There are actually several fits, we
  refer to the AAMQS paper~\cite{AAMQS} for a more detailed
  discussion.}.
This model {\it simultaneously} provides a good description of charged hadron
transverse momentum distributions in $p+p$ collisions at
7~TeV center of mass energy~\cite{pp_ptDistr}. This is a rather
non-trivial cross-check: the MV model initial
condition~(\ref{eq:N_MV}) overshoots the LHC data by roughly an
order of magnitude at $p_\perp\gsim 6$~GeV~\cite{pp_ptDistr}.

However, since the model~(\ref{eq:N_AAMQS}) was introduced
essentially ``by hand'' it is an open question how it extends to
nuclei. This is a crucial issue for predicting the nuclear
modification factor $R_{pA}$ for $p+Pb$ collisions at LHC, and for
heavy-ion structure functions which could be measured at a future
electron-ion eIC collider~\cite{eIC}. One possibility is that the
AAMQS modification of the MV model dipole is due to some unknown
$A$-independent non-perturbative effect. Here, we explore another
option, namely that for protons the effects due to the $\sim\rho^4$
operators may not be negligible.

\begin{figure}[htb]
\includegraphics[width=8cm]{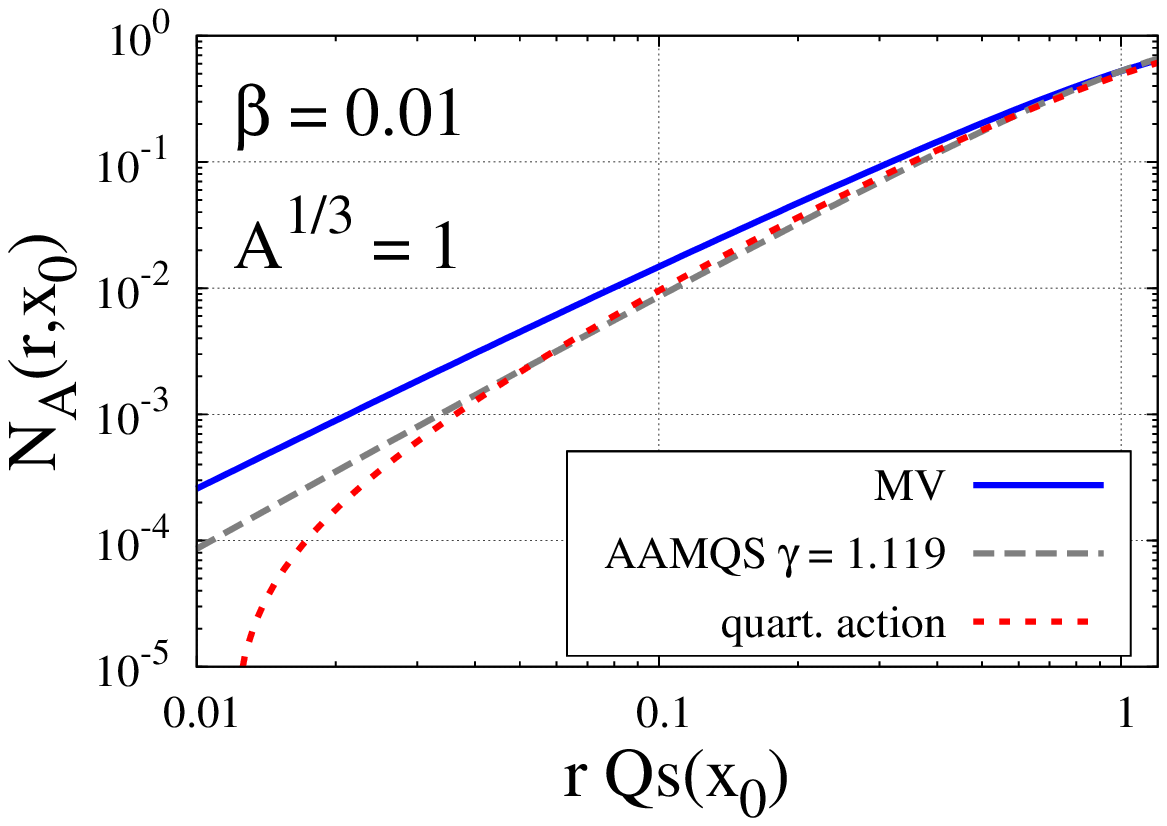}
\includegraphics[width=8cm]{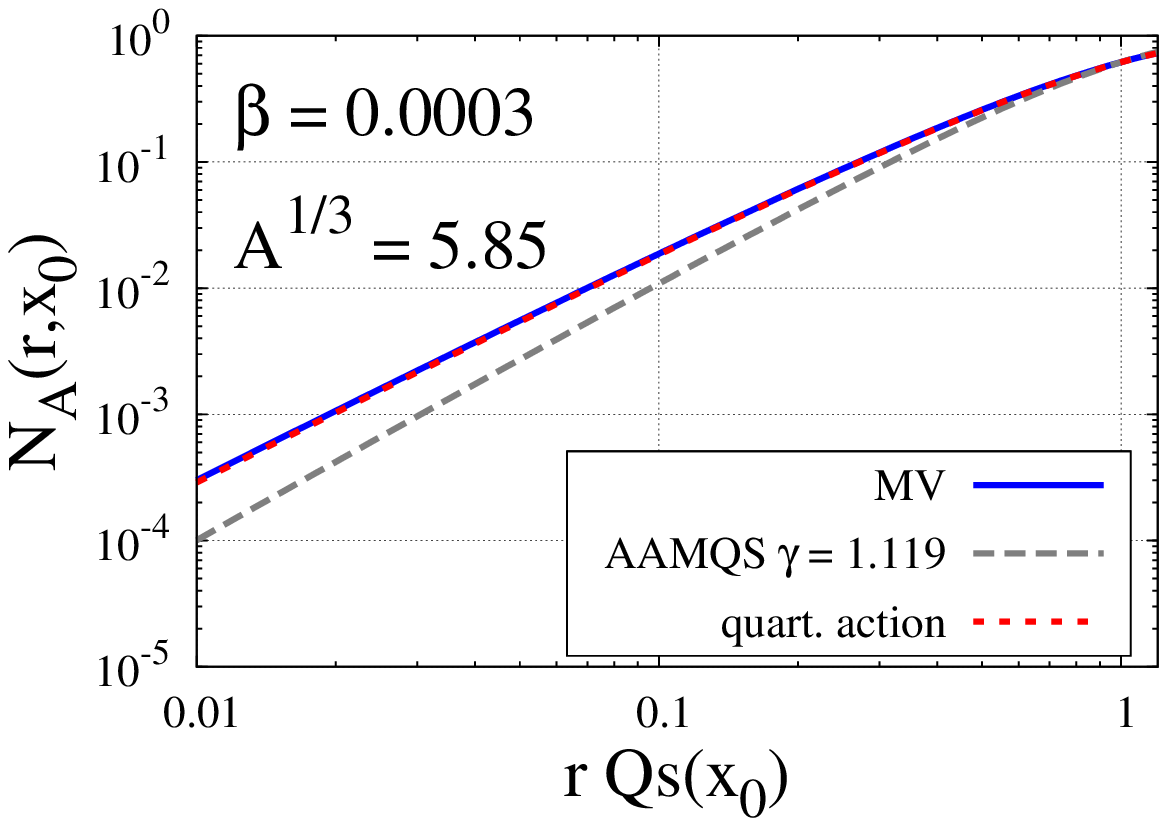}
\caption{Left: scattering amplitude for an adjoint dipole
  ($N_A=2N-N^2$) on a proton, assuming $Q_s^2=0.168$~GeV$^2$ and
  $\Lambda^2=0.0576$~GeV$^2$. 
  Right: same for a nucleus with $A=200$ and $Q_s^2\sim A^{1/3}$, 
  $\beta_A\sim A^{-2/3}$.}
\label{fig:QvsAAMQS}
\end{figure}

We can match our result~(\ref{eq:Nr_quartic}) approximately to the
AAMQS model by choosing
\be \label{eq:beta}
\beta \equiv 
\frac{C_F^2}{6\pi^3}\frac{g^8}{Q_s^2\, \kappa_4} 
\left[\int_{-\infty}^\infty dz^- \mu^4(z^-)\right]^2 \simeq
\frac{1}{100}~~~~,~~ (A=1).
\ee
For nuclei, $\beta_A\sim A^{-2/3}$ since each longitudinal integration over
$z^-$ is proportional to the thickness $\sim A^{1/3}$ of the nucleus
while $\mu^2(z^-)$ and $\mu^4(z^-)$ are $A$-independent. The
scattering amplitude for a dipole in the adjoint representation with
this $\beta$ is shown in fig.~\ref{fig:QvsAAMQS}.~\footnote{To
  regularize the behavior at large $r$, for this figure we have
  replaced $\log\frac{1}{r \Lambda} \to \log(e+\frac{1}{r \Lambda})$
  and assumed exponentiation of the $\mathcal{O}(r^2)$
  expression. This does not affect the behavior at $rQ_s<1$.}
One observes that the dipole scattering amplitude derived from the
quartic action is similar to the AAMQS model over a broad range,
$rQ_S\gsim0.04$. Discrepancies appear at very short distances where
none of the above can be trusted. A more careful and quantitative
matching of $\beta$ to the AAMQS fit beyond the leading $\log
1/r\Lambda\gg1$ approximation should be performed in the future.

On the right, we plot the scattering amplitude for a nucleus with
$A=200$ nucleons, assuming that $Q_s^2\sim A^{1/3}$ while $\beta_A\sim
A^{-2/3}$. This illustrates how expectation values obtained with the
quartic action converge to those from the MV model when the valence
charge density is high (i.e., at large $A^{1/3}$).
If our idea that the $\rho^4$ term in the action provides the
explanation for the AAMQS model is indeed correct then their
modification of the MV model should vanish like $\beta_A\sim
A^{-2/3}$. This should be observable via $R_{pA}$ at the LHC.

\begin{acknowledgments}
A.D.\ thanks M.~Gyulassy for lively and useful discussions during a
seminar at Columbia University. The diagrams shown in the appendix
have been drawn with JaxoDraw~\cite{jaxo} (\verb=http://jaxodraw.sourceforge.net/=).

We gratefully acknowledge support by the DOE Office of Nuclear Physics
through Grant No.\ DE-FG02-09ER41620, from the ``Lab Directed Research
and Development'' grant LDRD~10-043 (Brookhaven National Laboratory),
and for PSC-CUNY award 63382-0042, jointly funded by The Professional
Staff Congress and The City University of New York.
\end{acknowledgments}

\section{Appendix}

Expectation values of operators $O[\rho]$ are computed as
\be \label{eq:VEV}
\langle O[\rho] \rangle \equiv \int {\cal D}\rho \; O[\rho]\;
e^{-S[\rho]} ~\Big/ ~ \int {\cal D}\rho \;e^{-S[\rho]}~. \nonumber
\ee
We work perturbatively in 1/$\kappa_4$ and keep only
the first term in the expansion. Then,
\bea \label{eq:exp_value} 
\langle O[\rho]\rangle &\equiv& 
\frac{\int
\mathcal{D}\rho ~O[\rho]~e^{-S_G[\rho]}\left[1-\frac{1}{\kappa_4}\int
  d^2\bold v_\perp \int dv_1^-dv_2^-
  \rho^a_{v_1}\rho^a_{v_1}\rho^b_{v_2}\rho^b_{v_2}\right]}{\int
    \mathcal{D}\rho~ e^{-S_G[\rho]}\left[1-\frac{1}{\kappa_4}\int
      d^2\bold v_\perp \int dv_1^-dv_2^-
      \rho^a_{v_1}\rho^a_{v_1}\rho^b_{v_2}\rho^b_{v_2}\right]} \nonumber \\ 
&=& \frac{\left< O[\rho]\left(1-\frac{1}{\kappa_4}\int
    d^2\bold v_\perp \int dv_1^-dv_2^-
    \rho^a_{v_1}\rho^a_{v_1}\rho^b_{v_2}\rho^b_{v_2}\right)\right>_G}{\langle
    1-\frac{1}{\kappa_4}\int d^2\bold v_\perp \int dv_1^-dv_2^-
    \rho^a_{v_1}\rho^a_{v_1}\rho^b_{v_2}\rho^b_{v_2}\rangle_G}~~~.
  \eea
In lattice regularization the denominator evaluates to
\bea
& & \left<1-\frac{1}{\kappa_4}\int d^2\bold v_\perp \int dv_1^-dv_2^-
\rho^a_{v_1}\rho^a_{v_1}\rho^b_{v_2}\rho^b_{v_2}\right>_G\nonumber \\
&=& 
1-\frac{1}{\kappa_4}\frac{N_s}{\Delta \bold
  v_\perp}\left\{(N_c^2-1)^2\left[\int^\infty _{-\infty}
  dv^-\mu^2(v^-)\right]^2+2~(N_c^2-1)\int^\infty_{-\infty}
dv^-\mu^4(v^-)\right\}~~~,  \label{eq:Qnorm}
\eea
where $N_s$ denotes the number of lattice sites (the volume) and
$\Delta x_\perp$ the transverse area of a single lattice site (square
of lattice spacing). Also, we have used a local
$\langle\rho\rho\rangle$ correlation function:
\be \label{eq:2point}
\langle\rho^a(x^-,\bold x_\perp)\, \rho^b(y^-,\bold
y_\perp)\rangle=\delta^{ab}\mu^2(x^-)\delta(x^--y^-)\delta(\bold
x_\perp-\bold y_\perp)~.
\ee

\subsection{Dipole Operator}

We are interested in the expectation value of the dipole operator
defined as
\be
\hat{D}(\bold x_\perp,y_\perp)
\equiv\frac{1}{N_c}{\rm tr}~V(\bold x_\perp)V^\dagger(\bold y_\perp)~~~.
\ee
Here, $V$ denotes a Wilson line
\be
V(\bold x_\perp)= \mathcal{P}  \exp \left\{  -i g^2 \int_{-\infty}^\infty dz^-
\frac{1}{\nabla^2_\perp}\rho_a(z^-,\bold x_\perp)t^a \right\}~~~,
\ee
where
\be
\frac{1}{\nabla^2_\perp}\rho_a(z^-,\bold x_\perp)=\int d^2\bold
z_\perp G_0(\bold x_\perp-\bold z_\perp)\rho_a(z^-,\bold z_\perp)
= - \frac{1}{g}A^+~~~,
\ee
is proportional to the gauge potential in covariant gauge. The
matrices $t^a$ are the generators of the fundamental
representation of SU(3), normalized according to ${\rm tr}~t^a
t^b=\frac{1}{2}\delta^{ab}$.

$G_0$ is the static propagator which inverts the 2-dimensional Laplacian:
\be
\frac{\partial^2}{\partial\bold z^2_\perp}G_0(\bold x_\perp-\bold
z_\perp)=\delta(\bold x_\perp-\bold z_\perp)~~~;
\ee
\be \label{eq:popagator_def}
G_0(\bold x_\perp-\bold z_\perp)=
-\int\frac{d^2 \bold k_\perp}{(2\pi)^2}\frac{e^{i\bold k_\perp \cdot
    (\bold x_\perp-\bold z_\perp)}} {\bold k^2_\perp}~~~.
\ee
With this propagator we can write the Wilson line as
\be
V(\bold x_\perp)=\mathcal{P} \exp \left\{-ig^2\int^\infty_{-\infty} dz^-\int
d^2\bold z_\perp \, G_0(\bold x_\perp - \bold z_\perp)\, \rho_a(z^-,\bold
z_\perp) \, t^a \right\}~~~.
\ee
The correlator $\langle V(\bold x_\perp)V^\dagger(\bold
y_\perp)\rangle$ for a Gaussian (MV) action has already been
calculated before, see for example ref.~\cite{GelisPeshier}. The result is:
\be \label{eq:GaussianExpVal} \langle V(\bold x_\perp)V^\dagger(\bold
y_\perp)\rangle_G = \exp
\left\{-\frac{g^4}{2}(t^at_a)\left[\int_{-\infty}^\infty
  dz^-\mu^2(z^-)\right]\int d^2\bold z_\perp \left[G_0(\bold x_\perp
  -\bold z_\perp) -G_0(\bold y_\perp -\bold
  z_\perp)\right]^2\right\}~~.  
\ee
Note that this is diagonal in color (proportional to $1\!\!1_{3\times3}$).
To calculate the expectation value of the dipole operator with the new
action, we first expand the Wilson lines order by order in the
gauge coupling $g^2$,
\bea
V(\bold x_\perp)&=&1-ig^2\int d^2\bold z_{\perp 1}G_0(\bold x_\perp-\bold z_{\perp 1})\int^\infty_{-\infty} dz_1^-\rho^a(z_1)t^a\nonumber \\
&&-g^4\int d^2\bold z_{\perp 1}d^2\bold z_{\perp 2}G_0(\bold
x_\perp-\bold z_{\perp 1})G_0(\bold x_\perp-\bold z_{\perp
  2})\int^\infty_{-\infty} dz_1^-\int^\infty_{z_1^-}
dz_2^-\rho^a(z_1)\rho^b(z_2)t^at^b \nonumber \\
& & + \cdots
\eea
\bea
V^\dagger(\bold y_\perp)&=&1+ig^2\int d^2\bold u_{\perp 1}G_0(\bold y_\perp-\bold u_{\perp 1})\int^\infty_{-\infty} du_1^-\rho^a(u_1)t^a\nonumber \\
&&-g^4\int d^2\bold u_{\perp 1}d^2\bold u_{\perp 2}G_0(\bold
y_\perp-\bold u_{\perp 1})G_0(\bold y_\perp-\bold u_{\perp
  2})\int^\infty_{-\infty} du_1^-\int^{u_1^-}_{-\infty}
du_2^-\rho^a(u_1)\rho^b(u_2)t^at^b \nonumber\\
& & + \cdots
\eea
For brevity we only write the terms up to $\mathcal{O}(g^4)$ but below
we shall actually require terms up to $\mathcal{O}(g^8)$.

To zeroth order in $g$, the expectation value of the correlator is just 1:
\be
\mathcal O \left(g^0\right)=1~~~. \nonumber
\ee
The order $g^2$ contribution is zero because $\langle\rho^a(z)\rangle=0$:
\be
\mathcal O \left(g^2\right)=0~~~. \nonumber
\ee

\subsection{Order $g^4$}
The first non-trivial contribution arises at $\mathcal O (g^4)$ and is
given by the sum of expectation values of three terms (two from each
Wilson line and one mixed term):
\be
-g^4\int d^2\bold z_{\perp 1}d^2\bold z_{\perp 2}G_0(\bold x_\perp-\bold z_{\perp 1})G_0(\bold x_\perp-\bold z_{\perp 2})\int^\infty_{-\infty} dz_1^-\int^\infty_{z_1^-} dz_2^-\langle\rho^a(z_1)\rho^b(z_2)\rangle t^at^b~~,
\ee
\be
-g^4\int d^2\bold u_{\perp 1}d^2\bold u_{\perp 2}G_0(\bold y_\perp-\bold u_{\perp 1})G_0(\bold y_\perp-\bold u_{\perp 2})\int^\infty_{-\infty} du_1^-\int^{u_1^-}_{-\infty} du_2^-\langle\rho^a(u_1)\rho^b(u_2)\rangle t^at^b~~,
\ee
\be
g^4\int d^2\bold z_{\perp 1}\int d^2\bold u_{\perp 1}G_0(\bold x_\perp-\bold z_{\perp 1})G_0(\bold y_\perp-\bold u_{\perp 1})\int^\infty_{-\infty} dz_1^-\int^\infty_{-\infty} du_1^-\langle\rho^a(z_1)\rho^b(u_1)\rangle t^at^b~~.
\ee

\begin{figure}[htb]
\includegraphics[width=30mm]{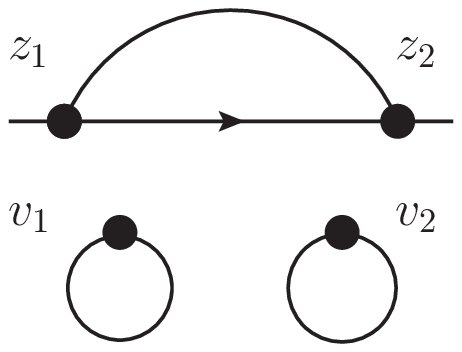}\hspace{1cm}
\includegraphics[width=30mm]{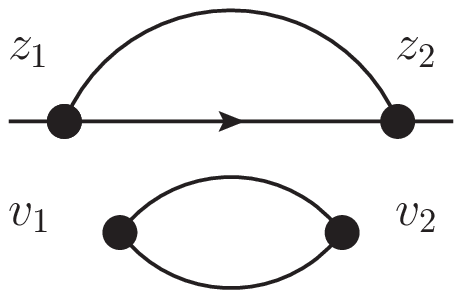}\hspace{1cm}
\includegraphics[width=30mm]{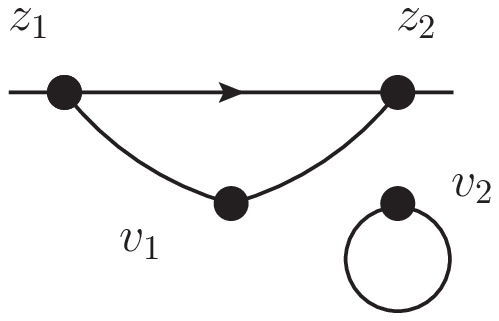}\hspace{1cm}
\includegraphics[width=30mm]{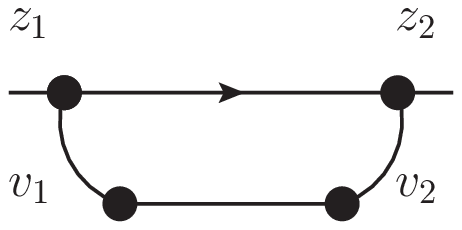}
\caption{$\langle V\rangle$ at order $g^4/\kappa_4$.}
\label{fig:g^4}
\end{figure}

Using~(\ref{eq:exp_value}) and~(\ref{eq:GaussianExpVal})
the first term becomes (we will divide by the
normalization factor later):
\bea 
&-& \frac{g^4}{2}t^at_a\int_{-\infty}^\infty dz^-\mu^2(z^-)\int
d^2\bold z_\perp G_0^2(\bold x_\perp-\bold z_\perp) \nonumber \\ 
&+ &
\frac{g^4}{\kappa_4}\int d^2\bold z_{\perp 1} d^2\bold z_{\perp 2}
d^2\bold v_\perp G_0(\bold x_\perp-\bold z_{\perp 1}) G_0(\bold
x_\perp-\bold z_{\perp 2}) \nonumber\\
& & \times  \int_{-\infty}^\infty dz_1^-
\int_{z_1^-}^\infty dz_2^- \int dv_1^- dv_2^- \langle
\rho_{z_1}^a\rho_{z_2}^b\rho_{v_1}^c\rho_{v_1}^c\rho_{v_2}^d\rho_{v_2}^d\rangle
t^a t^b ~.
\eea
All possible contractions for the second term in the above expression are
\bea
\langle\rho_{z_1}^a\rho_{z_2}^b\rho_{v_1}^c\rho_{v_1}^c\rho_{v_2}^d\rho_{v_2}^d\rangle=
\langle\rho_{z_1}^a\rho_{z_2}^b\rangle\left[\langle \rho_{v_1}^c\rho_{v_1}^c\rangle\langle \rho_{v_2}^d\rho_{v_2}^d\rangle+2~\langle\rho_{v_1}^c\rho_{v_2}^d\rangle\langle\rho_{v_1}^c\rho_{v_2}^d\rangle\right] \nonumber \\
+4~\langle\rho_{z_1}^a\rho_{v_1}^c\rangle\left[\langle\rho_{z_2}^b\rho_{v_1}^c\rangle \langle\rho_{v_2}^d\rho_{v_2}^d\rangle+2~\langle\rho_{z_2}^b\rho_{v_2}^d\rangle\langle\rho_{v_1}^c\rho_{v_2}^d\rangle\right],
\eea
and are shown diagrammatically in fig.~\ref{fig:g^4}.
Using~(\ref{eq:2point}) to perform the contractions and dividing also
by the
normalization factor~(\ref{eq:Qnorm}) leads us to\footnote{One has to
  be careful with performing the integration over $z_2^-$: $\int_{z_1^-}^\infty
dz_2^-\delta(z_1^--z_2^-)=1/2$. If $x^-$ is discretized, $z_2^-$
should be placed ahead of $z_1^-$ by at least half a lattice spacing
$\Delta x^-$. Similarly, when expanding a single Wilson lines to order
$g^6$: $z_3^- \ge z_2^-+\Delta x^-/2 \ge z_1^-+\Delta x^-$.}
\bea
 \frac{1}
{ 1-\frac{1}{\kappa_4}\frac{N_s}{\Delta \bold v_\perp}\left\{(N_c^2-1)^2\left[\int^\infty _{-\infty} dv^-\mu^2(v^-)\right]^2+2~(N_c^2-1)\int^\infty_{-\infty} dv^-\mu^4(v^-)\right\}}\times \nonumber\\
\left\{-\frac{g^4}{2}t^at_a\int_{-\infty}^\infty dz^-\mu^2(z^-)\int d^2\bold z_\perp G_0^2(\bold x_\perp- \bold z_\perp)\times\right.   \nonumber\\
\left. \left\{1-\frac{1}{\kappa_4}\frac{N_s}{\Delta \bold v_\perp}\left[(N_c^2-1)^2\left[\int^\infty _{-\infty} dv^-\mu^2(v^-)\right]^2+
2~(N_c^2-1)\int^\infty_{-\infty} dv^-\mu^4(v^-)\right]\right\}+\right. \nonumber \\
 \left.2\frac{g^4}{\kappa_4}\frac{t^at_a}{\Delta \bold v_\perp}\int d^2\bold z_\perp  G_0^2(\bold x_\perp-\bold z_\perp)\left[(N_c^2-1)\int_{-\infty}^\infty dz^-\mu^2(z^-)\int_{-\infty}^\infty dv^-\mu^4(v^-)+2\int_{-\infty}^\infty dz^-\mu^6(z^-)\right] 
\right\} ~~ .
\eea
The third line in the above expression cancels\footnote{This is the
  standard cancellation of disconnected diagrams.} the normalization
factor once the latter is expanded to leading order in $1/\kappa_4$ so
that the previous expression simplifies to
\bea 
&-& \frac{g^4}{2}t^at_a\int d^2\bold z_\perp G_0^2(\bold x_\perp-\bold
z_\perp) \nonumber \\
& & \times \left\{\int_{-\infty}^\infty
dz^-\mu^2(z^-) \right. \nonumber\\
& & \left. ~~ -\frac{4}{\kappa_4 \Delta \bold v_\perp}
\left[(N_c^2-1)\int_{-\infty}^\infty
  dz^-\mu^2(z^-)\int_{-\infty}^\infty
  dv^-\mu^4(v^-)+2\int_{-\infty}^\infty dz^-\mu^6(z^-)\right]\right\}~.
\eea
The correction is absorbed into a renormalization of
$\int_{-\infty}^\infty dz^-\mu^2(z^-)$ in order that the two-point
function $\langle\rho\rho\rangle$ remains unaffected by the quartic
term in the action:
\bea
& & \int_{-\infty}^\infty dz^-\tilde{\mu}^2(z^-) = \nonumber \\
& &
\int_{-\infty}^\infty dz^-\mu^2(z^-)
-\frac{4}{\kappa_4\Delta
  \bold v_\perp}\left[(N_c^2-1)\int_{-\infty}^\infty
  dz^-\mu^2(z^-)\int_{-\infty}^\infty
  dv^-\mu^4(v^-)+2\int_{-\infty}^\infty dz^-\mu^6(z^-)\right]~.
\eea
Finally, the expectation value of $V(x_\perp)$ to order $g^4$ can
simply be written as
\be \label{eq:g4_first _line}
-\frac{g^4}{2}t^at_a ~\int_{-\infty}^\infty dz^-\tilde{\mu}^2(z^-)\int
d^2\bold z_\perp G_0^2(\bold x_\perp-\bold z_\perp)~~.
\ee
Similarly, $\langle V^\dagger(y_\perp)\rangle$:
\be \label{eq:g4_second _line}
-\frac{g^4}{2}t^at_a ~\int_{-\infty}^\infty dz^-
\tilde{\mu}^2(z^-)\int d^2\bold u_\perp G_0^2(\bold y_\perp-\bold
u_\perp)~~.
\ee
The mixed term will be the same as the previous terms except with
a positive sign and without the factor of 1/2 which originated from
the path ordering ($z_1^-$ and $u_1^-$
are not ordered relative to each other):
\be \label{eq:g4_mix_term}
g^4t^at_a ~\int_{-\infty}^\infty dz^-\tilde{\mu}^2(z^-)\int d^2\bold
z_\perp G_0(\bold x_\perp-\bold z_\perp)G_0(\bold y_\perp-\bold
z_\perp)~~.
\ee
Summing~(\ref{eq:g4_first _line}), (\ref{eq:g4_second _line}) and
(\ref{eq:g4_mix_term}), we obtain the complete result at order $g^4$:
\be \label{eq:g4}
\mathcal O (g^4)=-\frac{g^4}{2}t^at_a~\int_{-\infty}^\infty
dz^-\tilde{\mu}^2(z^-)\int d^2\bold z_\perp \left[G_0(\bold
  x_\perp-\bold z_\perp)-G_0(\bold y_\perp-\bold z_\perp)\right]^2~~.
\ee
This is identical to the result obtained in the Gaussian theory once
the two-point function $\langle\rho\rho\rangle\sim\mu^2$ has been
matched. This was to be expected, of course, since only two-point
functions of $\rho$ arise at $\mathcal{O}(g^4)$. Note, also,
that~(\ref{eq:g4}) vanishes as $\bold y_\perp\to \bold x_\perp$.

\subsection{Order $g^8$}

Next, we consider order $g^8$. There are all in all five
terms: two of order $g^8$ from the expansion of a single Wilson line, two
mixed terms (order $g^2$ from the first line and $g^6$
from the second line and vice versa), and one term from
multiplying $g^4$ terms from both Wilson lines.

\subsubsection{$g^8$ from $V(\bold x_\perp)$}

\begin{figure}
  \centering
  \subfloat[~]{\label{fig:g^8VxDisConn}\includegraphics[width=30mm]{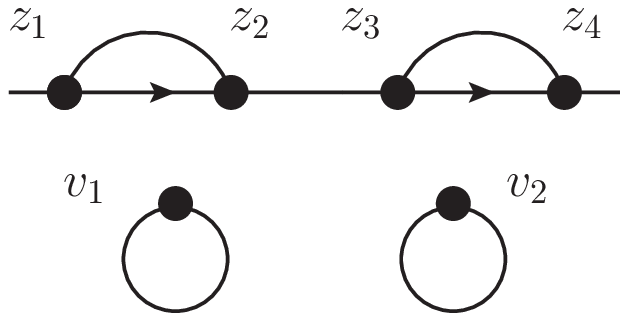}\includegraphics[width=30mm]{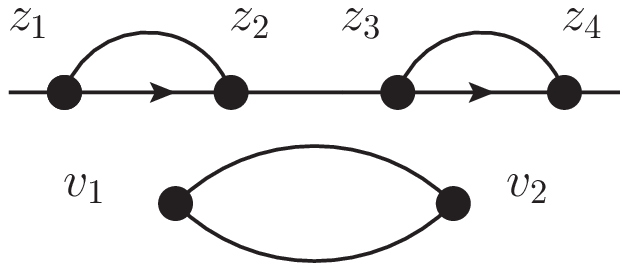}}                
  \subfloat[~]{\label{fig:g^8VxConn}\includegraphics[width=30mm]{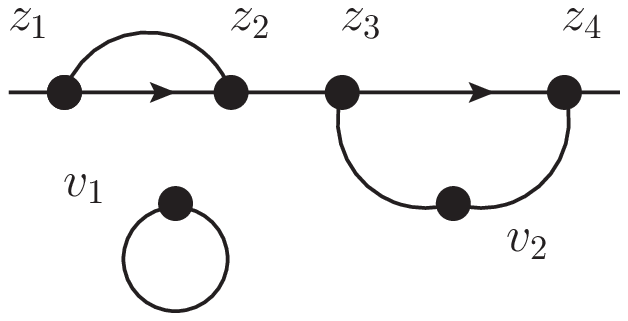}\includegraphics[width=30mm]{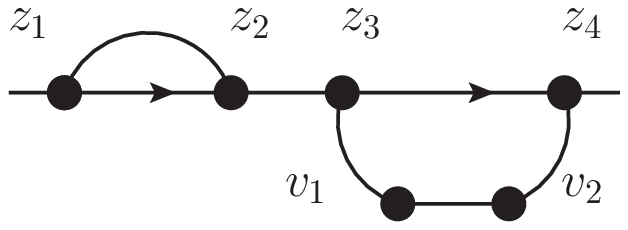}}
  \subfloat[~]{\label{fig:g^8VxCorr}\includegraphics[width=30mm]{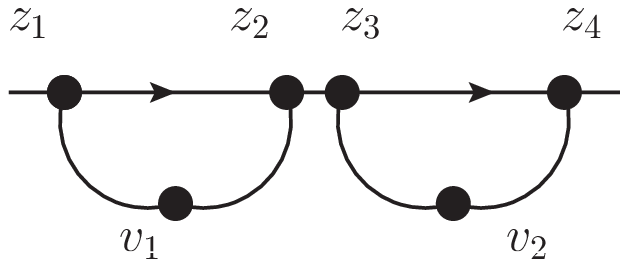}}
  \caption{Order $g^8/\kappa_4$ contribution to $\langle V\rangle$.}
  \label{fig:g^8Vx}
\end{figure}

First, we calculate $\langle V(\bold x_\perp)\rangle$ at order
$g^8$. In the Gaussian theory
\be
\frac{g^8}{8}\left(t^at_a\right)^2\left[\int^\infty_{-\infty}dz^-
\mu^2(z^-)\right]^2\left[\int
  d^2\bold z_\perp G_0^2(\bold x_\perp-\bold
  z_\perp)\right]^2~~. \nonumber
\ee
Again, using~(\ref{eq:exp_value}), the correction is
\bea
&-&\frac{g^8}{\kappa_4}\int d^2\bold z_{\perp 1} d^2\bold z_{\perp 2}
d^2\bold z_{\perp 3} d^2\bold z_{\perp 4} \int d^2\bold v_\perp
G_0(\bold x_\perp-\bold z_{\perp 1})G_0(\bold x_\perp-\bold z_{\perp
  2})G_0(\bold x_\perp-\bold z_{\perp 3})G_0(\bold x_\perp-\bold
z_{\perp 4}) \nonumber \\
& & \times
\int_{-\infty}^\infty dz_1^-\int_{z_1^-}^\infty
dz_2^-\int_{z_2^-}^\infty dz_3^-\int_{z_3^-}^\infty
dz_4^-\int_{-\infty}^\infty dv_1^-\int_{-\infty}^\infty dv_2^- \langle
\rho_{z_1}^a\rho_{z_2}^b \rho_{z_3}^c
\rho_{z_4}^d\rho_{v_1}^e\rho_{v_1}^e\rho_{v_2}^f\rho_{v_2}^f\rangle
t^a t^b t^c t^d~.
\eea
All possible contractions for $\langle \rho_{z_1}^a\rho_{z_2}^b
\rho_{z_3}^c
\rho_{z_4}^d\rho_{v_1}^e\rho_{v_1}^e\rho_{v_2}^f\rho_{v_2}^f\rangle$
are shown diagrammatically in fig.~\ref{fig:g^8Vx}.
The disconnected diagrams~\ref{fig:g^8VxDisConn} give the following
contribution:
\bea
&-& \frac{g^8}{8}\left(t^at_a\right)^2\left[\int_{-\infty}^\infty
  dz^-\mu^2(z^-)\right]^2\left[\int d^2\bold z_\perp G_0^2(\bold
  x_\perp-\bold z_\perp)\right]^2 \nonumber \\
& & \times
\frac{1}{\kappa_4}\frac{N_s}{\Delta \bold
  v_\perp}\left[(N_c^2-1)^2\left[\int^\infty _{-\infty}
    dv^-\mu^2(v^-)\right]^2+
2~(N_c^2-1)\int^\infty_{-\infty} dv^-\mu^4(v^-)\right]~,
\eea
which will cancel the normalization factor.

The connected diagrams from fig.~\ref{fig:g^8VxConn} again renormalize
$\mu^2(z^-)$ to $\tilde{\mu}^2(z^-)$:
\bea
\left[\int_{-\infty}^\infty dz^-\tilde{\mu}^2(z^-)\right]^2 &=&
\left[\int^\infty _{-\infty} dz^-\mu^2(z^-)\right]^2  \nonumber \\
& & \hspace{-3cm} -\frac{8}{\kappa_4\Delta \bold
  v_\perp}\left[(N_c^2-1)\left[\int_{-\infty}^\infty
    dz^-\mu^2(z^-)\right]^2 \int_{-\infty}^\infty
  dv^-\mu^4(v^-)+2\int_{-\infty}^\infty
  dz^-\mu^2(z^-)\int_{-\infty}^\infty dv^-\mu^6(v^-)\right]~.
\eea
Finally, the diagrams~\ref{fig:g^8VxCorr} give the correction
\be
-\frac{g^8}{\kappa_4}\left(t^at_a\right)^2 \left[\int_{-\infty}^\infty
  dz^-\mu^4(z^-)\right]^2\int d^2\bold z_\perp G_0^4(\bold
x_\perp-\bold z_\perp)~~.
\ee
That leads us to the complete $\mathcal O(g^8)$ contribution to
$\langle V(x_\perp)\rangle$:
\bea \label{eq:g^8_x}
& & \frac{g^8}{8}\left(t^at_a\right)^2~\left[\int_{-\infty}^\infty
  dz^-\tilde{\mu}^2(z^-)\right]^2~\left[\int d^2\bold z_\perp
  G_0^2(\bold x_\perp-\bold z_\perp)\right]^2\nonumber \\
&-& \frac{g^8}{\kappa_4}\left(t^at_a\right)^2 \left[\int_{-\infty}^\infty
  dz^-\mu^4(z^-)\right]^2\int d^2\bold z_\perp G_0^4(\bold
x_\perp-\bold z_\perp)~.
\eea
The expectation value of $V^\dagger(\bold y_\perp)$ is obtained from
the above by substituting $\bold x_\perp \to \bold y_\perp$.

\subsubsection{$g^6$ from $V(\bold x_\perp)$~$\times$~ $g^2$ from
  $V^\dagger(\bold y_\perp)$}

\begin{figure}
  \centering
  \subfloat[~]{\label{fig:g^6Vx_g^2VyDisConn}
\includegraphics[width=30mm]{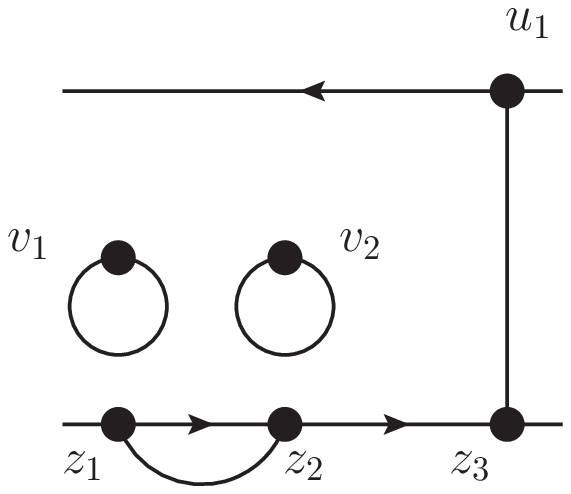}\includegraphics[width=30mm]{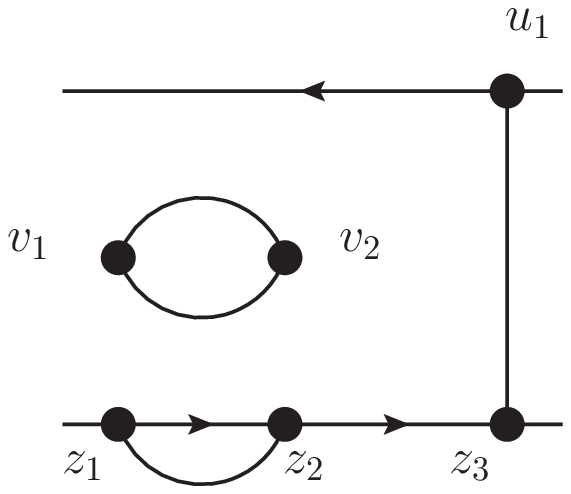}}

   \subfloat[~]{\label{fig:g^6Vx_g^2VyConn}\includegraphics[width=30mm]{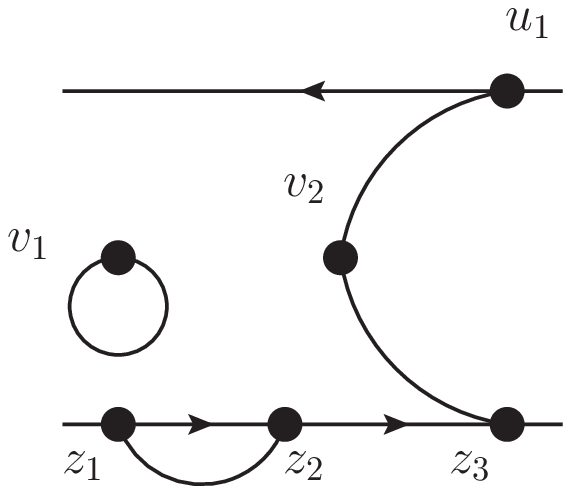}\includegraphics[width=30mm]{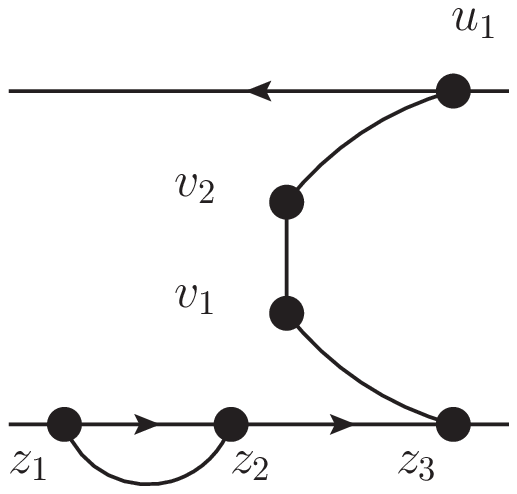}\includegraphics[width=30mm]{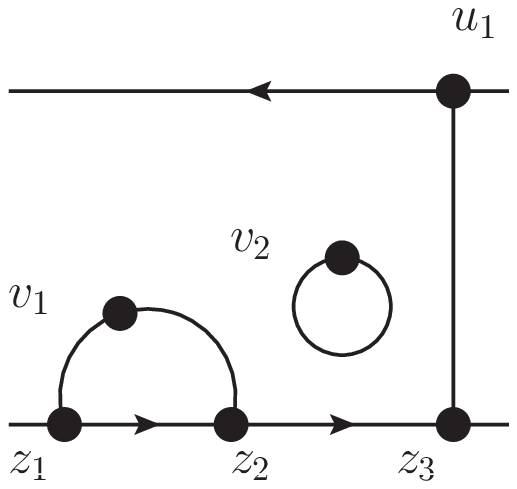}\includegraphics[width=30mm]{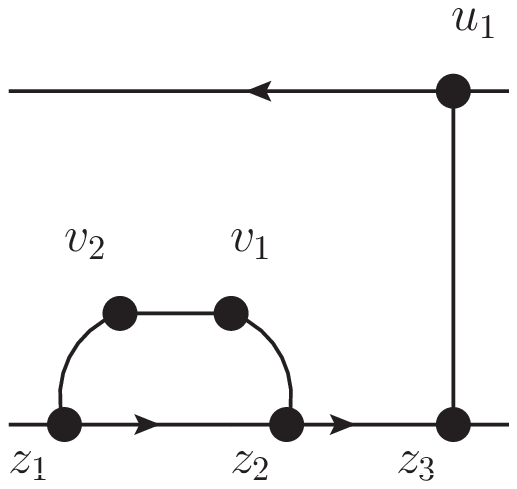}}

   \subfloat[~]{\label{fig:g^6Vx_g^2VyCorr}\includegraphics[width=30mm]{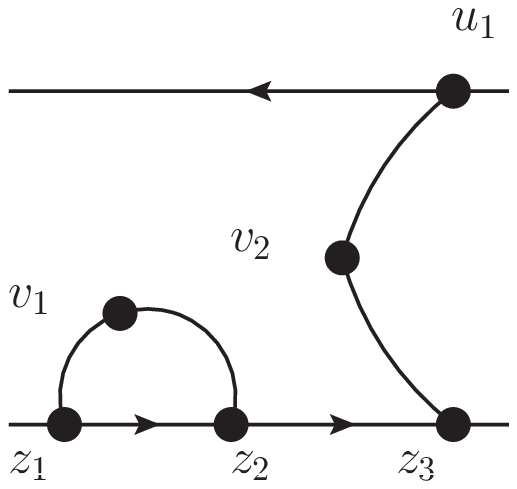}}
 \caption{$V(x_\perp) V^\dagger(y_\perp)$ at order $g^8/\kappa_4$ ($g^6$ from
   $V(\bold x_\perp)$~$\times$~ $g^2$ from $V^\dagger(\bold
   y_\perp)$).}
 \label{fig:g^6Vx_g^2Vy}
\end{figure}

Next is the mixed term obtained when multiplying the $g^6$ term from
the $x$ Wilson line with the $g^2$ term from the $y$ Wilson line. In
the Gaussian theory,
\be 
-\frac{g^8}{2}\left(t^at_a\right)^2 \left[\int_{-\infty}^\infty
dz^-\mu^2(z^-)\right]^2\int d^2\bold z_\perp G_0^2(\bold x_\perp-\bold
  z_\perp)\int d^2\bold u_\perp G_0(\bold x_\perp-\bold
  u_\perp)G_0(\bold y_\perp-\bold u_\perp)~.
\ee
The correction is:
\bea
&+& \frac{g^8}{\kappa_4}\int d^2\bold z_{\perp 1} d^2\bold z_{\perp 2}
d^2\bold z_{\perp 3}  G_0(\bold x_\perp-\bold z_{\perp 1})G_0(\bold
x_\perp-\bold z_{\perp 2})G_0(\bold x_\perp-\bold z_{\perp 3})\int
d^2\bold u_{\perp 1}G_0(\bold y_\perp-\bold u_{\perp 1}) \nonumber\\
& & \times \int d^2\bold v_\perp
\int_{-\infty}^\infty dz_1^-\int_{z_1^-}^\infty
dz_2^-\int_{z_2^-}^\infty dz_3^-\int_{-\infty}^\infty
du_1^-\int_{-\infty}^\infty dv_1^-\int_{-\infty}^\infty dv_2^-
\nonumber\\
& & \times \langle
\rho_{z_1}^a\rho_{z_2}^b \rho_{z_3}^c
\rho_{u_1}^d\rho_{v_1}^e\rho_{v_1}^e\rho_{v_2}^f\rho_{v_2}^f\rangle
t^a t^b t^c t^d~.
\eea
The possible contractions are given in fig.~\ref{fig:g^6Vx_g^2Vy}.
As before, the disconnected diagrams~\ref{fig:g^6Vx_g^2VyDisConn}
cancel against the normalization, while the diagrams
in~\ref{fig:g^6Vx_g^2VyConn} renormalize $\mu^2$ to
$\tilde{\mu}^2$; finally the
diagrams~\ref{fig:g^6Vx_g^2VyCorr} will give the correction. Summing
all these diagrams plus the Gaussian part, we get:
\bea \label{eq:g^6_x_X_g^2_y}
-\frac{g^8}{2}\left(t^at_a\right)^2 ~\left[\int_{-\infty}^\infty dz^-\tilde{\mu}^2(z^-)\right]^2~\int d^2\bold z_\perp G_0^2(\bold x_\perp-\bold z_\perp)\int d^2\bold u_\perp G_0(\bold x_\perp-\bold u_\perp)G_0(\bold y_\perp-\bold u_\perp) \nonumber \\
+4~\frac{g^8}{\kappa_4}\left(t^at_a\right)^2\left[\int_{-\infty}^\infty
  dz^-\mu^4(z^-)\right]^2\int d^2\bold z_\perp G_0^3(\bold
x_\perp-\bold z_\perp)G_0(\bold y_\perp-\bold z_\perp)~.
\eea
Once again, a similar contribution (with ${\bold x_\perp} \leftrightarrow
{\bold y_\perp}$) arises from $V(\bold x_\perp)$ at
$\mathcal{O}(g^2)$ times $V^\dagger(\bold y_\perp)$ at order $g^6$.

\subsubsection{$g^4$ from $V(\bold x_\perp)$~$\times$~ $g^4$ from
  $V^\dagger(\bold y_\perp)$}

The last term to consider is the one obtained when multiplying
$\mathcal O(g^4)$ from the $x$ Wilson line with $\mathcal O(g^4)$ from
the $y$ Wilson line. The Gaussian contribution is:
\bea 
\frac{g^8}{4}\left(t^at_a\right)^2 \left[\int_{-\infty}^\infty
dz^-\mu^2(z^-)\right]^2\int d^2\bold z_\perp G_0^2(\bold x_\perp-\bold
  z_\perp)\int d^2\bold u_\perp G_0^2(\bold y_\perp-\bold u_\perp)
  \nonumber \\ +\frac{g^8}{2}\left(t^at_a\right)^2
  \left[\int_{-\infty}^\infty dz^-\mu^2(z^-)\right]^2\left[\int
    d^2\bold z_\perp G_0(\bold x_\perp-\bold z_\perp) G_0(\bold
    y_\perp-\bold u_\perp)\right]^2~.
\eea
For the quartic action we have to add
\bea 
&-& \frac{g^8}{\kappa_4}\int d^2\bold z_{\perp 1} d^2\bold
z_{\perp 2} G_0(\bold x_\perp-\bold z_{\perp 1})G_0(\bold
x_\perp-\bold z_{\perp 2})\int d^2\bold u_{\perp 1} d^2\bold u_{\perp
  2}G_0(\bold y_\perp-\bold u_{\perp 1})G_0(\bold y_\perp-\bold
u_{\perp 2}) \nonumber\\
& & \times \int d^2\bold v_\perp
\int_{-\infty}^\infty dz_1^-\int_{z_1^-}^\infty
dz_2^-\int_{-\infty}^\infty du_1^-\int_{-\infty}^{u_1^-}
du_2^-\int_{-\infty}^\infty dv_1^-\int_{-\infty}^\infty dv_2^-\nonumber\\
& & \times \langle
\rho_{z_1}^a\rho_{z_2}^b \rho_{u_1}^c
\rho_{u_2}^d\rho_{v_1}^e\rho_{v_1}^e\rho_{v_2}^f\rho_{v_2}^f\rangle
t^a t^b t^c t^d~.
\eea
\begin{figure}
  \centering
  \subfloat[~]{\label{fig:g^4Vx_g^4VyDisConn}\includegraphics[width=30mm]{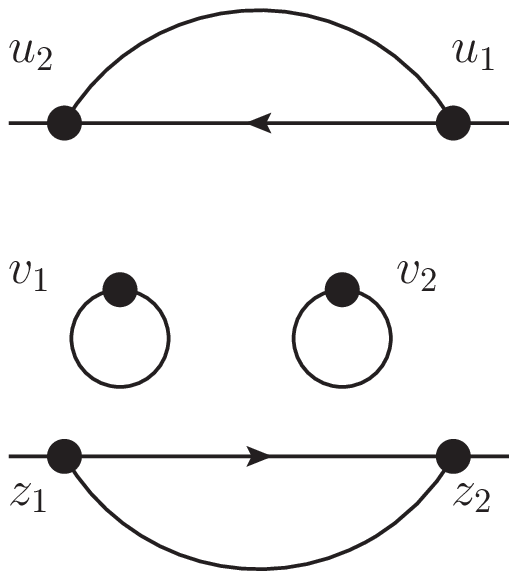}\includegraphics[width=30mm]{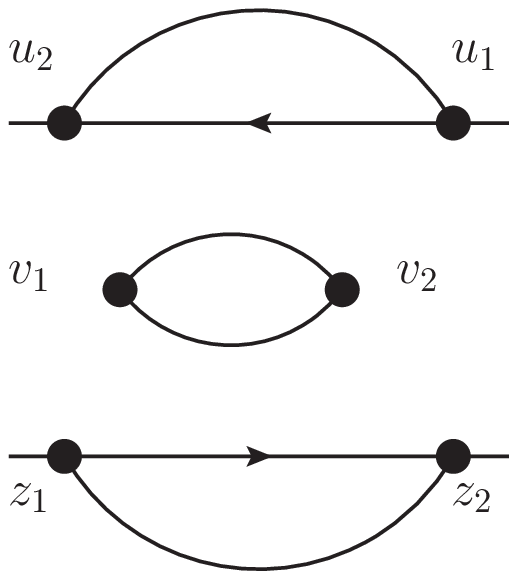}\includegraphics[width=30mm]{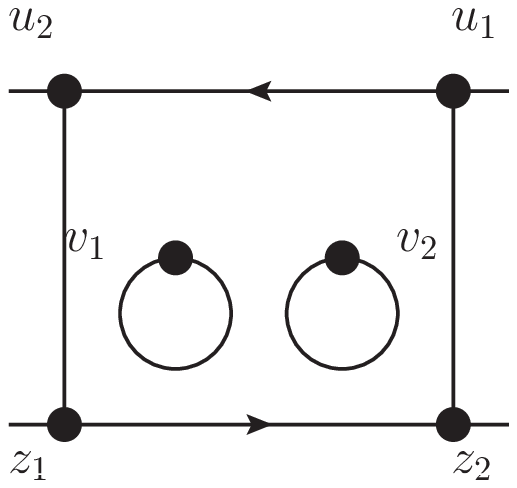}\includegraphics[width=30mm]{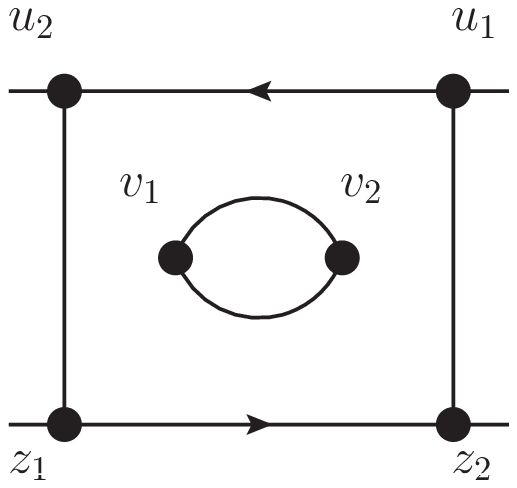}}                
  
   \subfloat[~]{\label{fig:g^4Vx_g^4VyConn}\includegraphics[width=30mm]{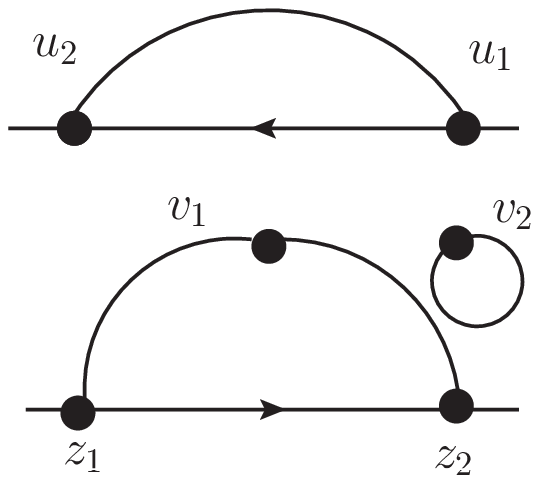}\includegraphics[width=30mm]{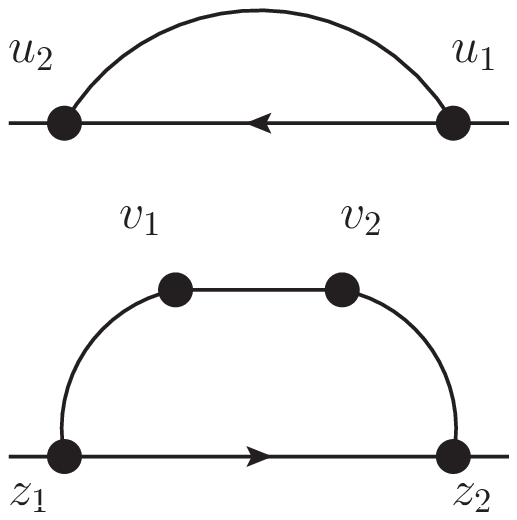}\includegraphics[width=30mm]{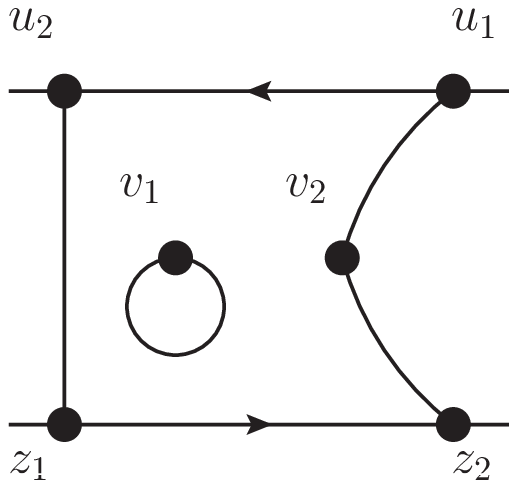}\includegraphics[width=30mm]{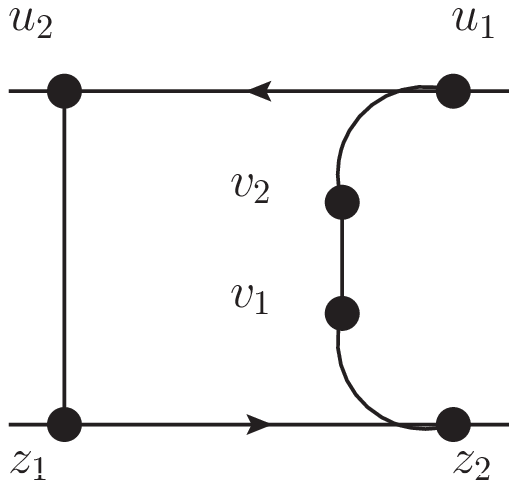}}

   \subfloat[~]{\label{fig:g^4Vx_g^4VyCorr}\includegraphics[width=30mm]{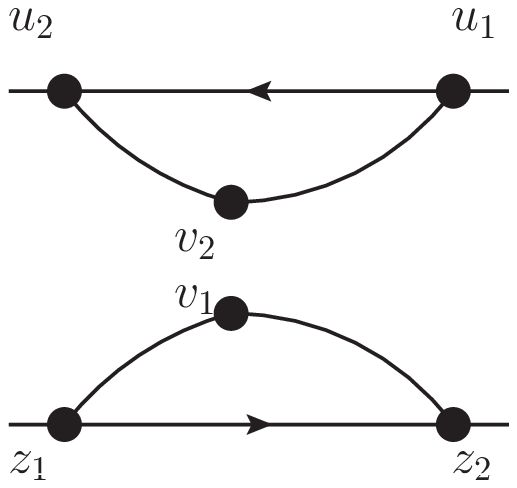}\includegraphics[width=30mm]{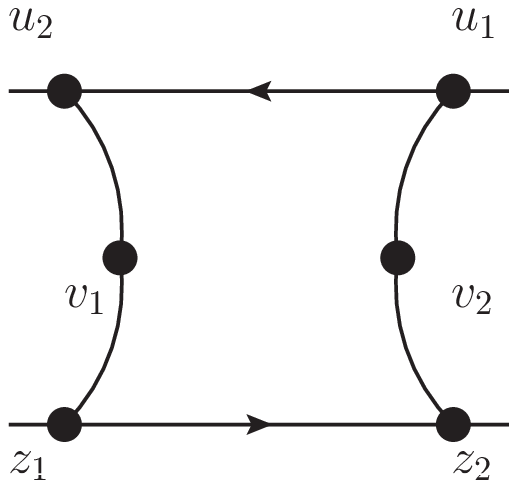}}
 \caption{Expectation value of
$V(\bold x_\perp)$ at order $g^4$ times $V(\bold y_\perp)$ at order $g^4$.}
 \label{fig:g^4Vx_g^4Vy}
\end{figure}
Following the same procedure as before, the diagrams from
fig.~\ref{fig:g^4Vx_g^4Vy} give:
\bea \label{eq:g^4_x_X_g^4_y} \frac{g^8}{4}\left(t^at_a\right)^2~
\left[\int_{-\infty}^\infty dz^-\tilde{\mu}^2(z^-)\right]^2~\int
d^2\bold z_\perp G_0^2(\bold x_\perp-\bold z_\perp)\int d^2\bold
u_\perp G_0^2(\bold y_\perp-\bold u_\perp) \nonumber \\ +
\frac{g^8}{2}\left(t^at_a\right)^2~ \left[\int_{-\infty}^\infty
  dz^-\tilde{\mu}^2(z^-)\right]^2~ \left[\int d^2\bold z_\perp
  G_0(\bold x_\perp-\bold z_\perp) G_0(\bold y_\perp-\bold
  u_\perp)\right]^2 \nonumber
\\ -6~\frac{g^8}{\kappa_4}\left(t^at_a\right)^2
\left[\int_{-\infty}^\infty dz^-\mu^4(z^-)\right]^2\int d^2\bold
z_\perp G_0^2(\bold x_\perp-\bold z_\perp) G_0^2(\bold y_\perp-\bold
z_\perp)~.  \eea

\subsubsection{Complete order $g^8$}
Combining eqs.~(\ref{eq:g^8_x}+$x_\perp\leftrightarrow y_\perp$), 
(\ref{eq:g^6_x_X_g^2_y}+$x_\perp\leftrightarrow y_\perp$),
and~(\ref{eq:g^4_x_X_g^4_y}) we get the complete contribution to the
expectation value of the dipole at order $g^8$:
\bea \frac{1}{2}\left(\frac{g^4\left(t^at_a\right)}{2}\right)^2~
\left[\int_{-\infty}^\infty dz^-\tilde{\mu}^2(z^-)\right]^2~\left[\int
  d^2\bold z_\perp\left[G_0(\bold x_\perp-\bold z_\perp)-G_0(\bold
    y_\perp-\bold z_\perp)\right]^2\right]^2 \nonumber
\\ -\frac{g^8}{\kappa_4}\left(t^at_a\right)^2
\left[\int_{-\infty}^\infty dz^-\mu^4(z^-)\right]^2 \int d^2\bold
z_\perp\left[G_0(\bold x_\perp-\bold z_\perp)-G_0(\bold y_\perp-\bold
  z_\perp)\right]^4~. \nonumber \eea
Note that this again vanishes as $\bold y_\perp \to \bold x_\perp$.

\subsection{Complete expectation value of the dipole operator}

Adding together the terms of order 1, $g^4$ and $g^8$, the expectation
value of the dipole operator becomes
\bea
D(x_\perp-y_\perp) &\equiv& \langle\frac{1}{N_c}{\rm tr}~V(\bold
x_\perp)V^\dagger(\bold y_\perp)\rangle = \nonumber \\ 
1 &-& \frac{g^4}{2}C_F~\int_{-\infty}^\infty dz^-\tilde{\mu}^2(z^-)\int
d^2\bold z_\perp \left[G_0(\bold x_\perp-\bold z_\perp)-G_0(\bold
  y_\perp-\bold z_\perp)\right]^2 \nonumber \\
&+&
\frac{1}{2}\left(\frac{g^4 C_F}{2}\right)^2~
\left[\int_{-\infty}^\infty dz^-\tilde{\mu}^2(z^-)\right]^2~\left[\int
  d^2\bold z_\perp\left[G_0(\bold x_\perp-\bold z_\perp)-G_0(\bold
    y_\perp-\bold z_\perp)\right]^2\right]^2 \nonumber \\
&-& \frac{g^8}{\kappa_4} C_F^2 \left[\int_{-\infty}^\infty dz^- \mu^4(z^-)\right]^2 \int d^2\bold z_\perp\left[G_0(\bold x_\perp-\bold z_\perp)-G_0(\bold y_\perp-\bold z_\perp)\right]^4+\cdots
\eea
where 
\be
C_F=\frac{N_c^2-1}{2N_c}~. \nonumber
\ee

To write this in a more compact form we introduce the saturation scale
\be
{Q}_s^2\equiv \frac{g^4}{2\pi}C_F \int_{-\infty}^\infty dz^-\tilde{\mu}^2(z^-)~,
\ee
so that
\bea
D(r) &=& 1 - \pi Q_s^2 \int d^2\bold z_\perp 
  \left[G_0(\bold x_\perp-\bold z_\perp)-G_0(\bold
  y_\perp-\bold z_\perp)\right]^2 \nonumber \\
&+&
\frac{\pi^2}{2} Q_s^4 \left[\int
  d^2\bold z_\perp\left[G_0(\bold x_\perp-\bold z_\perp)-G_0(\bold
    y_\perp-\bold z_\perp)\right]^2\right]^2 \nonumber \\
 & -&
\frac{g^8}{\kappa_4} C_F^2 \left[\int_{-\infty}^\infty dz^-
  \mu^4(z^-)\right]^2 \int d^2\bold z_\perp\left[G_0(\bold
  x_\perp-\bold z_\perp)-G_0(\bold y_\perp-\bold z_\perp)\right]^4~.
\label{eq:Dr49}
\eea
Here, $r=|\bold x_\perp - \bold y_\perp|$.

\subsection{Explicit evaluation of $D(r)$ to leading $\log 1/r$ accuracy}

It is useful to obtain an explicit expression for $D(r)$ in the limit
$\log 1/r \Lambda \gg 1$, where $\Lambda$ is an infrared cutoff on
the order of the inverse nucleon radius.

The first non-trivial term in eq.~(\ref{eq:Dr49}) gives
\bea
& & \pi Q_s^2 \int d^2\bold z_\perp 
  \left[G_0(\bold x_\perp-\bold z_\perp)-G_0(\bold
  y_\perp-\bold z_\perp)\right]^2 \nonumber \\
&=& Q_s^2 \int\limits_0^\infty \frac{dk}{k^3} \left[1- J_0(kr)\right]
\simeq \frac{1}{4} r^2 Q_s^2 \log \frac{1}{r\Lambda}~,
\eea
in the leading $\log\, 1/r\Lambda \gg 1$ approximation.

Next, we need to compute the integral
\be
\int d^2\bold z_\perp\left[G_0(\bold x_\perp-\bold z_\perp)-G_0(\bold
  y_\perp-\bold z_\perp)\right]^4~.
\ee
From eq.~(\ref{eq:popagator_def}) for the propagator, 
\bea
\int d^2 \bold z_\perp G_0^4(\bold x_\perp -\bold z_\perp) &=&
\frac{1}{(2 \pi)^8}\int d^2\bold z_\perp \int d^2
\bold{k}_1 d^2 \bold{k}_2 d^2 \bold{k}_3 d^2
\bold{k}_4 \frac{1}{\bold{k}^2_1 \bold{k}^2_2
  \bold{k}^2_3 \bold{k}^2_4} e^{i\left(\bold{k}_1
  +\bold{k}_2 +\bold{k}_3+\bold{k}_4\right) \cdot
  (\bold x_\perp-\bold z_\perp)} \\
& & \hspace{-2cm} =
\frac{1}{(2 \pi)^6}\int d^2\bold K d^2 \bold k d^2 \bold Q d^2 \bold q
\frac{\delta (\bold K +\bold Q)}{\left(\frac{\bold K}{2}+\bold
  k\right)^2 \left(\frac{\bold
    K}{2}-\bold k\right)^2 \left(\frac{\bold Q}{2}+\bold q\right)^2
  \left(\frac{\bold Q}{2}-\bold q\right)^2} \\ 
&=& \frac{1}{(2
  \pi)^6}\int d^2\bold K \left[\int d^2 \bold k
  \frac{1}{\left(\frac{\bold K}{2}+\bold k\right)^2 \left(\frac{\bold
      K}{2}-\bold k\right)^2}\right]^2 ~. 
\eea
We regularize the integral in the square brackets by introducing a
cutoff $\Lambda$:
\be
\int d^2 \bold k \frac{1}{\left(\left(\frac{\bold K}{2}+\bold
k\right)^2+\Lambda^2\right) \left(\left(\frac{\bold K}{2}-\bold
k\right)^2+\Lambda^2\right)} =
\frac{2\pi}{\bold K^2}\log\frac{\bold K^2}{\Lambda^2}~.
\ee
Then,
\bea
\int d^2 \bold z_\perp G_0^4(\bold x_\perp -\bold z_\perp)=
\frac{1}{(2 \pi)^4}\int  \frac{d^2\bold K}{\bold K^4}\log^2\frac
     {\bold K^2}{\Lambda^2} \nonumber \\
=\frac{1}{(2\pi)^3}\int_\Lambda^\infty \frac{d \bold K}{K^3}
\log^2\frac {\bold K^2}{\Lambda^2} = \frac{1}{(2\pi)^3} \frac
    {1}{\Lambda^2}~.
\eea
Following the same procedure for $\int d^2 \bold z_\perp G_0^2(\bold
x_\perp -\bold z_\perp)G_0^2(\bold y_\perp -\bold z_\perp)$ we arrive
at:
\bea \int d^2 \bold z_\perp G_0^2(\bold x_\perp -\bold
z_\perp)G_0^2(\bold y_\perp -\bold z_\perp)=\frac{1}{(2 \pi)^4}\int
\frac{d^2\bold K}{\bold K^4} e^{i \bold K \cdot (\bold x_\perp-\bold
  y_\perp)} \log^2\frac {\bold K^2}{\Lambda^2} \nonumber \\ \nonumber
\\ =\frac{1}{(2\pi)^3}\int_\Lambda^\infty \frac{d^2 \bold K}{
  K^3} J_0(\bold K r)\log^2\frac {\bold
  K^2}{\Lambda^2}=\frac{1}{(2\pi)^3} \left(
\frac{1}{\Lambda^2}+\frac{1}{3}r^2\log^3(r\Lambda)\right)+\mathcal
O\left[r^2\log^2(r\Lambda)\right]~. \nonumber \eea
Similarly,
\bea \int d^2 \bold z_\perp G_0^3(\bold x_\perp -\bold
z_\perp)G_0(\bold y_\perp -\bold z_\perp)= \nonumber \\ \frac{1}{(2
  \pi)^5}\int \frac{d^2\bold K}{\bold K^2} e^{i \frac{\bold K}{2}
  \cdot (\bold x_\perp-\bold y_\perp)} \log\frac {\bold
  K^2}{\Lambda^2} \int d^2 \bold q \frac{e^{i\bold q \cdot(\bold
    x_\perp -\bold y_\perp)}}{\left(\frac{\bold K}{2}+\bold q\right)^2
  \left(\frac{\bold K}{2}-\bold q\right)^2} ~.
\eea
Using: 
\bea 
& & \int d^2 \bold q \frac{e^{i\bold q \cdot(\bold x_\perp -\bold
y_\perp)}}{\left(\frac{\bold K}{2}+\bold q\right)^2 \left(\frac{\bold
        K}{2}-\bold q\right)^2} \nonumber \\ 
&\approx& \int d^2 \bold q
    \frac{1+i\bold q\cdot \bold r}{\left(\frac{\bold K}{2}+\bold
      q\right)^2 \left(\frac{\bold K}{2}-\bold
      q\right)^2}=\frac{2\pi}{\bold K^2}\log\frac{\bold
      K^2}{\Lambda^2}+\mathcal O(Kr)~,
\eea
we get
\bea \int d^2 \bold z_\perp G_0^3(\bold x_\perp -\bold
z_\perp)G_0(\bold y_\perp -\bold z_\perp) \nonumber \\ =\frac{1}{(2
  \pi)^3} \int_\Lambda^\infty \frac{d^2 \bold K}{K^3}
J_0\left(\frac{1}{2} K r\right)\log^2\frac {\bold
  K^2}{\Lambda^2}=\frac{1}{4}\frac{1}{(2
  \pi)^3}\left(\frac{4}{\Lambda^2}+\frac{1}{3}r^2 \log^3(r\Lambda)\right)~,
\eea
so that, finally,
\bea
 D(r) &=&
1 - \frac{r^2 Q_s^2}{4}\log\frac{1}{r\Lambda}
+ \frac{1}{6\pi^3}\frac{g^8}{\kappa_4} C_F^2
  \left[\int_{-\infty}^\infty dz^-\mu^4(z^-)\right]^2
  r^2\log^3\frac{1}{r\Lambda} \\
&=& 1 - \frac{r^2 Q_s^2}{4}\log\frac{1}{r\Lambda}
+ \beta \,r^2 Q_s^2 \, \log^3\frac{1}{r\Lambda}~.
\eea
$\beta$ has been defined in eq.~(\ref{eq:beta}).

Performing a Fourier transform we obtain the transverse momentum
dependence of the dipole for $k_\perp\gg Q_s \gg\Lambda$:
\bea
 D(k_\perp) &\approx& 2\pi \frac{Q_s^2}{k_\perp^4}-
 \frac{g^8}{\pi^2 \kappa_4} C_F^2
  \left[\int_{-\infty}^\infty dz^-\mu^4(z^-)\right]^2 
 \frac{1}{k_\perp^4}{\log^2\frac{k_\perp^2}{\Lambda^2}} \\
&=& 2\pi \frac{Q_s^2}{k_\perp^4}- 6\pi\beta\,
 \frac{Q_s^2}{k_\perp^4} \, \log^2\frac{k_\perp^2}{\Lambda^2} ~.
\eea
The first term was taken from appendix~B of ref.~\cite{GelisPeshier}.
The second term provides a correction to the classical {\em
  bremsstrahlung} tail for finite valence parton density.

\subsection{Gluon density}

One can define~\cite{hep-ph/9802440} a Weizs\"acker-Williams like
gluon density of a nucleus, in the 
limit of small $\bold x_\perp^2$, as
\be \label{eq:xG}
x G_A(x,\bold x_\perp^2) = \frac{\left(N_c^2-1\right)\pi
  R^2}{\pi^2\alpha_s N_c}\frac{1}{\bold x_\perp^2}N(\bold x_\perp^2)~.
\ee
For the quartic action
\bea \label{eq:r_perp} 
N(x_\perp^2)= \frac{Q_s^2 x_\perp^2}{4}\log\frac{1}{x_\perp
  \Lambda}-\beta x_\perp^2 Q_s^2 \log^3\frac{1}{x_\perp \Lambda}~.
\eea
The first term, via eq.~(\ref{eq:xG}), gives $xG_A\sim
A$~\cite{hep-ph/9802440}. The second term gives a correction
\be 
-\frac{\left(N_c^2-1\right)\pi R^2}{\pi^2 \alpha_s N_c}
\beta \; Q_s^2
\log^3\frac{1}{x_\perp \Lambda}~\sim~-A^\frac{1}{3}~.  
\ee

\subsection{Form of the quartic term in the action}

In this section we explain why the $\rho$'s in the quartic term of the
action should sit at two different points in the longitudinal
direction in order that $N(r)$ vanishes as $\sim r^2$ as required by
color transparency.

\begin{figure}
  \centering
\includegraphics[width=38mm]{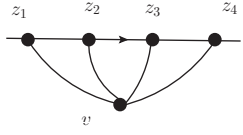}
  \caption{$\langle V\rangle$ at order $g^8/\kappa_4$ for the
    action~(\ref{eq:Ssamex-}).}
  \label{fig:g^8Vx_corr}
\end{figure}

First, let us note that the correction to the dipole scattering
amplitude due to the quartic term in the action results from
the diagrams~\ref{fig:g^8VxCorr}, ~\ref{fig:g^6Vx_g^2VyCorr} and
~\ref{fig:g^4Vx_g^4VyCorr}.
Now, let us consider the action:
\bea
S[\rho] &=& \int d^2\bold v_\perp\int^\infty_{-\infty} dv^- \left\{
\frac{\rho^a(v^-,\bold v_\perp) \rho^a(v^-,\bold
  v_\perp)}{2\mu^2(v^-)} \right.\nonumber \\
& & \left. +\frac{\rho^a(v^-,\bold v_\perp)\rho^a(v^-,\bold
  v_\perp)\rho^b(v^-,\bold v_\perp)\rho^b(v^-,\bold
  v_\perp)}{\kappa_4}\right\}~.
\label{eq:Ssamex-}
\eea
The analogue of diagram~\ref{fig:g^8VxCorr} for $\langle V\rangle$ at
order $g^8$ for this action is shown in
fig.~\ref{fig:g^8Vx_corr}. However, this diagram vanishes due to the
longitudinal path ordering in the Wilson line.

\begin{figure}
  \centering
\includegraphics[width=32mm]{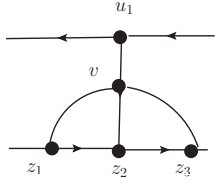}
  \caption{Order $g^6$ from $V(\bold x_\perp)$ times order $g^2$
    from $V^\dagger(\bold y_\perp)$ at order $1/\kappa_4$.}
  \label{fig:g6Vx_g2Vy_corr}
\end{figure}

The same reasoning applies to the analogue of ~\ref{fig:g^6Vx_g^2VyCorr}
shown in fig.~\ref{fig:g6Vx_g2Vy_corr}. In this case the three points
$z_1^-$, $z_2^-$ and $z_3^-$, can not be connected simultaniously to
one $v^-$ point. The delta functions coming from this kind of
contractions, $\delta(z_1^--v^-)\delta(z_2^--v^-)\delta(z_3^--v^-)$,
imply overlap of $z_1^-$ and $z_3^-$ which is not there in a path
ordered integral.

\begin{figure}
  \centering
\includegraphics[width=32mm]{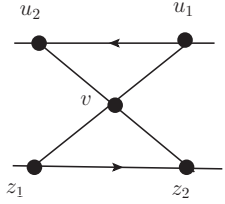}
  \caption{Order $g^4$ from $V(\bold x_\perp)$ times order $g^4$ from
    $V^\dagger(\bold y_\perp)$ at order $1/\kappa_4$.}
  \label{fig:g4Vx_g4Vy_corr}
\end{figure}
That leaves us with only one type of diagram proportional to
$1/\kappa_4$, shown in fig.~\ref{fig:g4Vx_g4Vy_corr}. This diagram is
not zero since there is no relative ordering between the points $z^-$
and $u^-$ on the two lines. This diagram has a constant
$r$-independent contribution which does not cancel because of the
missing diagrams~\ref{fig:g^8Vx_corr}
and~\ref{fig:g6Vx_g2Vy_corr}. This is (only terms at order $1/\kappa_4$
are given)
\bea 
D(r) &\sim& 
 1-\frac{3g^8}{\kappa_4}(C_F^2-\frac{2}{N_c})\int^\infty_{-\infty}
  dz^-\mu^8(z^-)\int d^2 \bold z_\perp G_0^2(\bold x_\perp-\bold
  z_\perp)G_0^2(\bold y_\perp-\bold z_\perp) \nonumber  \\ 
&=& 1 - \frac{3g^8}{(2\pi)^3\kappa_4}(C_F^2-\frac{2}{N_c})\int^\infty_{-\infty}
  dz^-\mu^8(z^-)\left(\frac{1}{\Lambda^2}-\frac{1}{3} r^2
  \log^3\left(\frac{1}{r \Lambda}\right)\right)~.
\eea
Hence we see that for the action~(\ref{eq:Ssamex-}) that $N(r)=1-D(r)$
approaches a constant as $r\to0$, in violation of color transparency.

For numerical (lattice gauge) computations of expectation
values~(\ref{eq:VEV}) it may be easier to integrate $\rho(x^-)$ and to
drop the longitudinal path ordering,
\bea
\tilde\rho(\bold x_\perp) &\equiv& \int\limits_{-\infty}^\infty dx^-
\rho(x^-,\bold x_\perp) \\
V(\bold x_\perp) &=& e^{i g^2 \frac{1}{\nabla_\perp^2}
\tilde\rho(\bold x_\perp)} ~.
\eea
For a detailed discussion, we refer to ref.~\cite{Lappi:2007ku}.
One would then consider the two-dimensional action
\be
S[\tilde\rho]= \int d^2\bold x_\perp 
\left\{ \frac{\tilde\rho^a(\bold x_\perp) \tilde\rho^a(\bold x_\perp)}{2\mu^2}
+ \frac{\tilde\rho^a(\bold x_\perp)\tilde\rho^a(\bold x_\perp)\tilde\rho^b(\bold
  x_\perp)\tilde\rho^b(\bold x_\perp)}{\kappa_4}\right\}~.
\ee
The diagrams from figs.~\ref{fig:g^8Vx_corr}
and~\ref{fig:g6Vx_g2Vy_corr} then do exist and cancel the
$r$-independent contribution from fig.~\ref{fig:g4Vx_g4Vy_corr} so
that again $N(r)\sim r^2$ at $r\to0$.


\end{document}